\documentclass[journal]{vgtc}                     

\onlineid{0}

\vgtccategory{Research}

\vgtcpapertype{please specify}

\title{SASAV: \underline{S}elf-Directed \underline{A}gent for \underline{S}cientific \underline{A}nalysis and \underline{V}isualization}

\author{%
  \authororcid{Jianxin Sun}{0000-0002-9627-9397},
  \authororcid{David Lenz}{0000-0002-2587-2783}, 
  \authororcid{Tom Peterka}{0000-0002-0525-3205}, and
  \authororcid{Hongfeng Yu}{0000-0002-0596-8227}
}

\authorfooter{
  \item
  	Jianxin Sun and Hongfeng Yu are with the University of Nebraska-Lincoln.
  	E-mail: s-jsun5@unl.edu, yu@cse.unl.edu.
  \item
  	David Lenz and Tom Peterka are with Argonne National Laboratory.
  	E-mail: dlenz@anl.gov, tpeterka@mcs.anl.gov.
}

\abstract{%
With recent advances in frontier multimodal large language models (MLLMs) for data understanding and visual reasoning, the role of LLMs has evolved from passive LLM-as-an-interface to proactive LLM-as-a-judge, enabling deeper integration into the scientific data analysis and visualization pipelines. However, existing scientific visualization agents still rely on domain experts to provide prior knowledge for specific datasets or visualization-oriented objective functions to guide the workflow through iterative feedback. This reactive, data-dependent, human-in-the-loop (HITL) paradigm is time-consuming and does not scale effectively to large-scale scientific data. In this work, we propose a Self-Directed Agent for Scientific Analysis and Visualization (SASAV), the first fully autonomous AI agent to perform scientific data analysis and generate insightful visualizations without any external prompting or HITL feedback. SASAV is a multi-agent system that automatically orchestrates data exploration workflows through our proposed components, including automated data profiling, context-aware knowledge retrieval, and reasoning-driven visualization parameter exploration, while supporting downstream interactive visualization tasks. This work establishes a foundational building block for the future AI for Science to accelerate scientific discovery and innovation at scale.
}


\teaser{
  \centering
  \includegraphics[trim=2 2 2 2,clip,width=\linewidth, alt={A view of a city with buildings peeking out of the clouds.}]{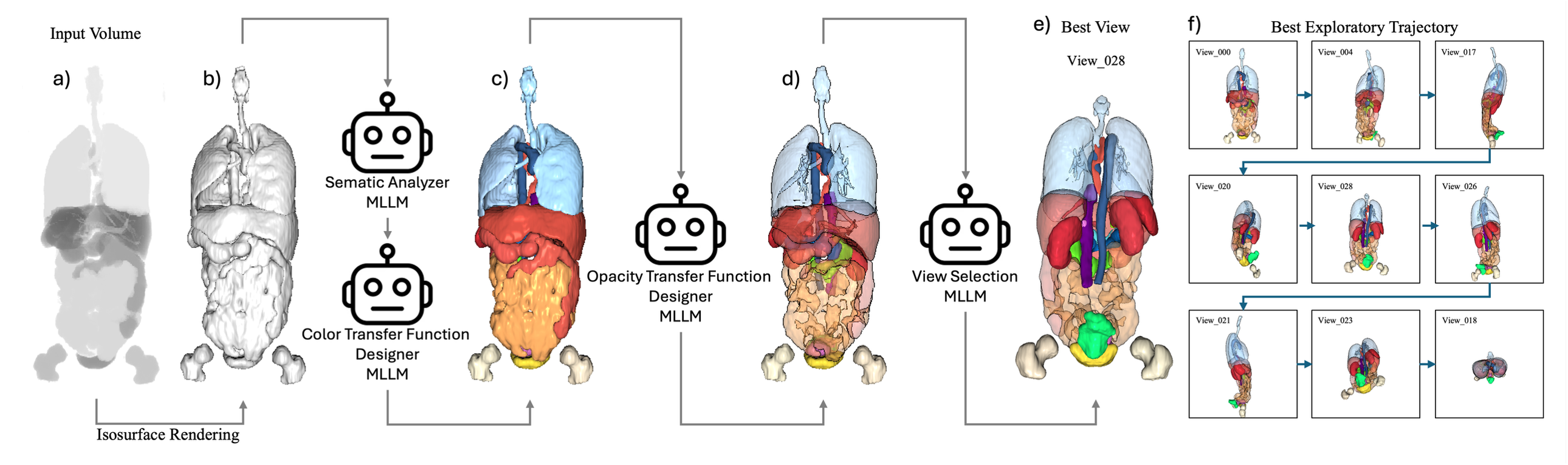}
  \caption{SASAV workflow to automatically generate visualization results for AbdomenAtlas 1.0 Mini datasets: (a) Direct volume rendering of the input volume; (b) Isosurface rendering of evenly sampled isovalues; (c) Isosurface rendering with color mapping suggested by the semantic analyzer and color transfer function designer; (d) Isosurface rendering with opacity mapping suggested by the opacity transfer function designer; (e) Isosurface rendering from the best viewpoint; (f) The animation rendering from the best viewpoint trajectory. Both the best view and the best exploratory trajectory are suggested by the view selection.}
  \label{fig:teaser}
}

\graphicspath{{figs/}{figures/}{pictures/}{images/}{./}} 

\usepackage{tabu}                      
\usepackage{booktabs}                  
\usepackage{lipsum}                    
\usepackage{mwe}                       

\usepackage{mathptmx}                  

\usepackage{adjustbox}
\usepackage{amsmath}
\usepackage{algorithm}
\usepackage{algpseudocode}
\usepackage{multirow}
\usepackage{pifont}
\usepackage[x11names]{xcolor}
\begin{document}




\firstsection{Introduction}
\maketitle



Scientific data analysis involves processing and interpreting raw data to extract meaningful patterns, relationships, and insights, while scientific visualization transforms that data into visual representations that humans can easily understand. Scientific visualization is a critical analytical tool that enables domain scientists to uncover meaningful features in complex datasets across a wide range of fields, including medical imaging, geophysics, meteorology, materials science, and physical simulations. Traditional scientific visualization systems are grounded in human-centered design to facilitate data exploration and support interactive workflows, leveraging techniques such as visual encoding, dimensionality reduction, level-of-detail (LOD) control, feature extraction, and transfer function design~\cite{10.1016/j.cosrev.2025.100731, 8740868}. Recent agentic AI systems demonstrate strong potential for intuitive interaction, data understanding, and visual reasoning by leveraging rapidly advancing frontier multimodal large language models (MLLMs)~\cite{11153807}.


\begin{figure}[t]
    \centering
        \includegraphics[trim=0 0 0 0,clip,width=\linewidth]{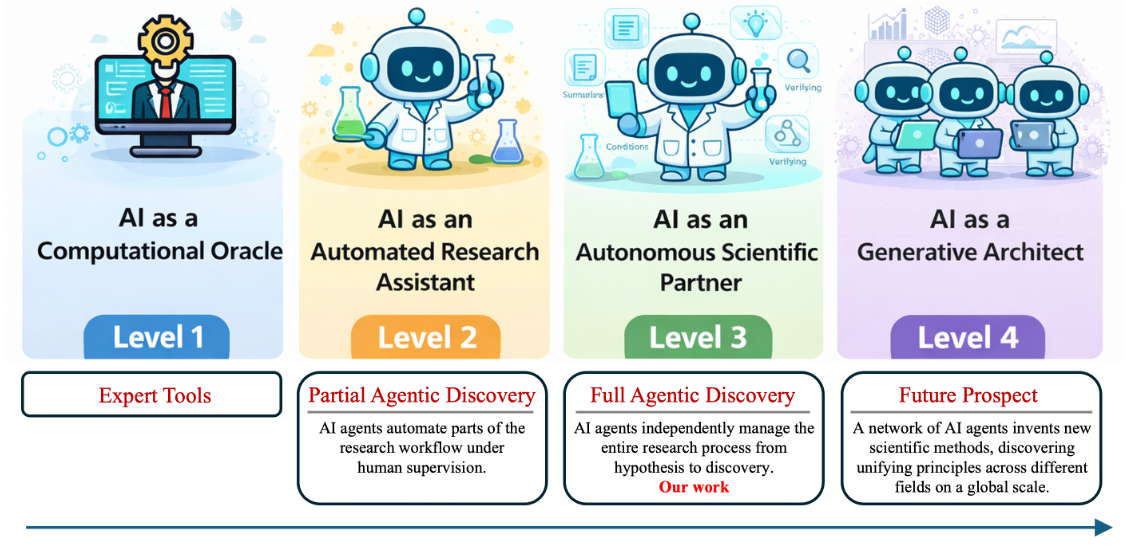}
    \caption{Evolution of AI for Science}
    \label{fig:ai4science}
\end{figure}

Existing Scientific Visualization (SciVis) agents basically function as reactive tools, relying on human-in-the-loop (HITL) input to provide prior knowledge of the target dataset and heuristic to direct data exploration. SciVis code generation LLM assistants, such as ChatVis\cite{11298870}, rely on users to provide a complete and specific description of visualization parameters, including the viewpoints, transfer functions (TFs), lighting, and visualization method, in order to converge through an iterative image metric-based self-correction. For interactive visualization of scientific data, LLM-based visualization approaches, such as NLI4VolVis~\cite{11264350} and ParaView-MCP~\cite{liu2025paraview}, require a clear visualization task description and continuous feedback from domain expert to guide the data exploration. Those iterative trial-and-error feedback loops with HITL is time-consuming, subjective, and infeasible when handling large-scale scientific data~\cite{pauloski2025empowering} with extensive visualization parameters.

MLLMs have demonstrated strong capabilities in image understanding and analysis, with perceptual abilities that closely align with human visual assessment~\cite{10.5555/3692070.3692324, 10670425, 10656242}. Recent work on visualization parameter optimization seeks to automate the search for key visualization parameters, such as the color and opacity transfer function, by employing an LLM-as-a-judge~\cite{GU2026101253} to provide feedback on a predefined objective function, leveraging the visual reasoning capability of the frontier MLLMs. AVA~\cite{https://doi.org/10.1111/cgf.15093} provides an automatic adjustment of the opacity TF according to the feedback from MLLMs, which evaluate the recognizability and clarity of intermediate renderings. Similarly, IntuiTF~\cite{wang2025intuiTF} introduces a framework that leverages MLLMs to define a loss metric for downstream evolutionary optimization, enabling the automatic exploration and evaluation of transfer functions for volume rendering. However, those methods suffer from the following issues: 1) Both AVA and IntuiTF remain semi-autonomous, as they still depend on users to specify clear visualization objectives: AVA requires a textual prompt, such as action instructions or visual perception prompts, while IntuiTF requires user text or visual intent. 2) There are two types of scientific data: empirical data and simulated data. Empirical data are scientific data measured or reconstructed from real-world objects, originating from physical reality with rich semantic meaning. Simulated data are scientific data generated by computational models that simulate physical, chemical, or biological processes. Existing methods are more effective for empirical data, as it is easier for users to provide semantically clear prompts to MLLMs, whereas simulated data is often difficult to interpret due to its blurry, distribution-like appearance. 3) Existing TF refinement is carried out in an extremely large and high-dimensional search space through a single-threaded iterative loop that queries MLLMs for ad hoc evaluations~\cite{11298870, liu2025paraview, 11264350}, leading to slow convergence and limited generalizability. 4) The visualization–perception–action loop adopted in existing methods tends to converge to local optima, leading to incomplete visualizations that miss important regions of interest and compromise the information richness~\cite{https://doi.org/10.1111/cgf.12934}. All the aforementioned issues limit the usability and scalability of MLLM-based agentic workflow for scientific visualization tasks, particularly when dealing with real-world, large-scale scientific data.




In this work, we propose SASAV, a self-directed agent for scientific analysis and visualization, which, to the best of our knowledge, is the first fully autonomous and zero-prompting agentic workflow for scientific visualization. SASAV provides a unified solution for visualizing various types of scientific data, where prior knowledge is not available to provide a specific task description. Our solution leverages the visual reasoning and comparison capabilities of the frontier MLLMs, enabling comprehensive, systematic, and efficient parameter optimization for visualization. We employ a multi-agent workflow with parallel and batched MLLM inference to enable efficient exploration of visualization parameters. SASAV consists of 4 key workflow components (data profiling, knowledge retrieval, TFs suggestion, and view selection) and directly generates 3 visualization forms (static visualization image, animation, and interactive visualization) to best help domain experts to accelerate scientific discovery. Our work is not designed to replace existing interactive visualizations, but rather to provide a comprehensive and informative starting point that facilitates more efficient exploration, ultimately accelerating scientific discovery and innovation. We see our work as one of the foundational building blocks for establishing the level 3 (fully agentic discovery) stage during the evolution of AI for Science~\cite{wei2025ai}, which is shown in \cref{fig:ai4science}. The demonstration, evaluation, and source code are available at \url{https://selfdirectedscivisagent.github.io}. The main contributions of the proposed fully autonomous self-directed AI agent for scientific data analysis and visualization include:
\setlist{nolistsep}
\begin{itemize}[leftmargin=*]
  \item A fully autonomous agentic workflow for large-scale scientific data analysis and visualization without any prior knowledge and HITL.
  \item A knowledge retrieval component to suggest regions of interest with scientific significance to guide the visualization.
  \item Efficient optimization of visualization parameters for color and opacity recommendation using a parallel multi-agent MLLM pipeline across the full range of the input data.
  \item View selection to determine the best viewpoint and view trajectory for informative static and animated visualizations.
\end{itemize}


\section{Related Work} 
\label{related work}

\subsection{Natural Language for Visualization}
Generative AI (GenAI) has been transforming data visualization in the past decade~\cite{YE202443}. In recent years, the Large Language Model (LLM) has been integrated into a visualization framework, demonstrating textual generative capacity and reasoning for efficient interaction through natural languages. Natural Language to Visualization (NL2VIS)~\cite{10.1007/s42979-025-04636-4} refers to a class of systems that automatically generate data visualizations from natural language queries. Shuai et al. presented DeepVIS~\cite{11284783}, an NL2VIS framework that leverages popular prompting strategies for step-by-step reasoning to improve the accuracy and transparency of generating data visualizations from natural language. Ouyang et al. proposed NVAGENT~\cite{ouyang-etal-2025-nvagent}, a collaborative agent workflow for NL2VIS to convert natural language queries into accurate data visualizations. Tian et al. proposed ChartGPT~\cite{10443572} to use a fine-tuned LLM with step-by-step reasoning to generate data visualization charts from abstract natural-language queries. Similarly, Maddigan et al. proposed Chat2VIS~\cite{10121440} to use LLMs to convert natural language queries directly into code that generates data visualizations. Guo et al. implemented Talk2Data~\cite{10.1145/3643894}, which enables exploratory data analysis by decomposing complex natural-language questions into simpler ones and answering them with visualizations. Zhao et al. proposed LEVA~\cite{10458347} to integrate large language models into visual analytics to guide exploration and automatically summarize insights. NL2VIS benchmarks are also proposed to evaluate the LLM's visual understanding and generation capability. Chen et al. provided VisEval~\cite{10670425}, a large-scale benchmark and automated evaluation framework for evaluating how well large language models generate visualizations from natural language queries. Luo et al. proposed nvBench 2.0~\cite{luo2025nvbench} to evaluate text-to-visualization systems on ambiguous queries by providing multiple valid visualizations and reasoning paths. Existing NL2VIS work focuses on the understanding and evaluation of chart-based visualization. Natural language for scientific visualization, NL4SicVis, is challenging due to its large-scale data size, in situ data generation, high-dimensionality, and time-varying nature.

\subsection{LLM-based Scientific Visualization Agents}
In recent years, preliminary studies have explored the use of natural language as an intuitive interface for describing scientific visualization tasks and delivering feedback, enabling iterative refinement of visualization outcomes. Peterka et al. introduced a way of doing "Vibe Coding" for visualization tasks by proposing ChatVis~\cite{11298870}, a retrieval-augmented and self-correcting LLM assistant that improves the accuracy and reliability of generating scientific visualization code from user intent through natural language. While early efforts focused on natural-language-driven interfaces for visualization, subsequent work has progressed beyond code generation to develop an agentic workflow that can directly orchestrate complex visualization systems. Liu et al. proposed ParaView-MCP, a visualization agent that integrates MLLMs with ParaView~\cite{ahrens2005paraview} via the Model Context Protocol (MCP)~\cite{10.1145/3796519} to enable natural-language-driven, vision-guided, and interactive control of scientific visualization workflows, thereby lowering the barrier to using complex tools. GMX-VMD-MCP is an open-standard interface that allows AI agents to directly execute, visualize, and analyze molecular dynamics simulations by connecting LLMs to the GROMACS~\cite{https://doi.org/10.1002/jcc.20291} and Visual Molecular Dynamics (VMD)~\cite{HUMPHREY199633} software suites. While these systems emphasize integration with established visualization and simulation tools, more recent approaches further extend this paradigm by enabling real-time, semantically rich interaction and direct manipulation of volumetric data. Ai et al. proposed NLI4VolVis~\cite{11264350}, a unified system that combines multi-agent LLMs, vision-language models, and editable 3D Gaussian splatting to enable real-time, semantic, natural-language-driven exploration and editing of volumetric data. However, existing LLM-based scientific visualization agents are largely passive, relying on explicit human-defined task descriptions to support visualization generation, feature highlighting, and parameter optimization.

\begin{figure*}[t]
    \centering 
    \includegraphics[trim=2 5 2 2,clip,width=\textwidth]{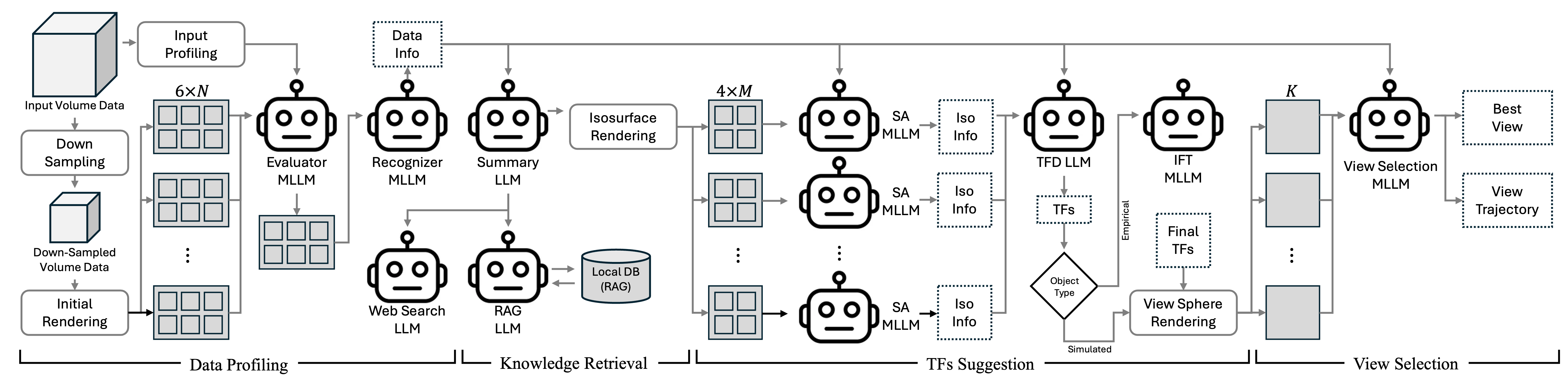}
    \caption{Architecture of SASAV. $N$ is the number of RSV selected for initial rendering, $M$ is the number of isovalues selected for isosurface rendering, and $K$ is the number of viewpoints sampled on the view surfaces.}
    \label{fig:arch}
\end{figure*}

\subsection{LLM-driven Autonomous Data Visualization}
Traditional automation in data visualization utilizes user-defined metrics or heuristics to guide the user to find the optimal visualization parameters, like transfer function design~\cite{4906854} and view selection~\cite{1532833}. Recent LLMs have created new possibilities to automate the data analysis and visualization process, leveraging the perceiving and reasoning capability of the foundation models. Dibia ~\cite{dibia-2023-lida} proposed LIDA, an LLM-based system that automatically generates data visualizations and infographics from datasets. Zhao et al. proposed ProactiveVA~\cite{11300819}, an LLM-based UI agent that proactively assists users during visual data analysis by detecting their needs and providing context-aware guidance. Wang et al. proposed Data Formulator~\cite{10292609} to automatically transform data to generate visualizations from high-level user concepts through an AI agent. Shen et al. proposed Data Director~\cite{10766492}, an LLM-based multi-agent system that automatically converts raw data into an animated visualization. Wang et al. introduced InsightLens~\cite{10989518}, an LLM-agent-based interactive system that automatically extracts, organizes, and visualizes insights and supporting evidence for more efficient data exploration. However, current proactive autonomous methods focus mainly on information visualization and still depend on interactive user input to guide the LLM. Liu et al. proposed AVA~\cite{https://doi.org/10.1111/cgf.15093}, a scientific visualization solution that converges to a predefined objective by combining natural language understanding with visual perception through iterative refinement. Wang et al. proposed IntuiTF~\cite{wang2025intuiTF}, a framework that uses MLLMs and evolutionary optimization to automatically explore and evaluate transfer functions for volume rendering based on user intent. However, both AVA and IntuiTF are semi-autonomous because it still requires users to provide clear visualization-related objectives. AVA requires a textual prompt, such as action instructions or visual perception prompts, whereas IntuiTF requires user text or visual intent. Due to the rendering overhead of iterative searching for optimal transfer functions, current scientific visualization agents are still limited to small static volumetric datasets. A fully autonomous AI agent is still missing in visualizing scientific data, addressing the specific challenges from the large-scale, high-dimensional, and time-varying nature of the scientific data.

\section{Methods}
\cref{fig:arch} illustrates the architecture of the primary workflow in SASAV, which consists of four steps: 1) Data profiling identifies the high-level characteristics of the input scientific data, including data format, underlying scientific object, and initial viewpoint, through consecutive queries of the proposed evaluator and recognizer. 2) Knowledge retrieval suggests regions of interest with high scientific significance from the internet and the local database to guide visualization emphasis and highlight key features. 3) Transfer function suggestion provides both color and opacity mappings by leveraging MLLM-based visual reasoning across isosurface renderings from different value levels. 4) View selection provides suggestions for the best viewpoint and the best exploratory trajectory by leveraging the summary and planning capability of the MLLM. SASAV employs a multi-agent architecture with parallel MLLM inference to enable an efficient and scalable workflow. In the paper, we constrain our scope to static volumetric scalar fields, as the proposed approach can be readily extended to time-varying scalar or vector fields. For visualization, we focus on Direct Volume Rendering (DVR) and isosurface rendering, as they are among the most widely adopted techniques for scientific data visualization and serve as effective starting points for downstream visualization tasks. We consider only 1D transfer functions in this work, as they provide a global visualization that is well-suited as an initial starting point.

\begin{figure}[t]
    \centering
        \includegraphics[trim=2 2 2 2,clip,width=0.9\linewidth]{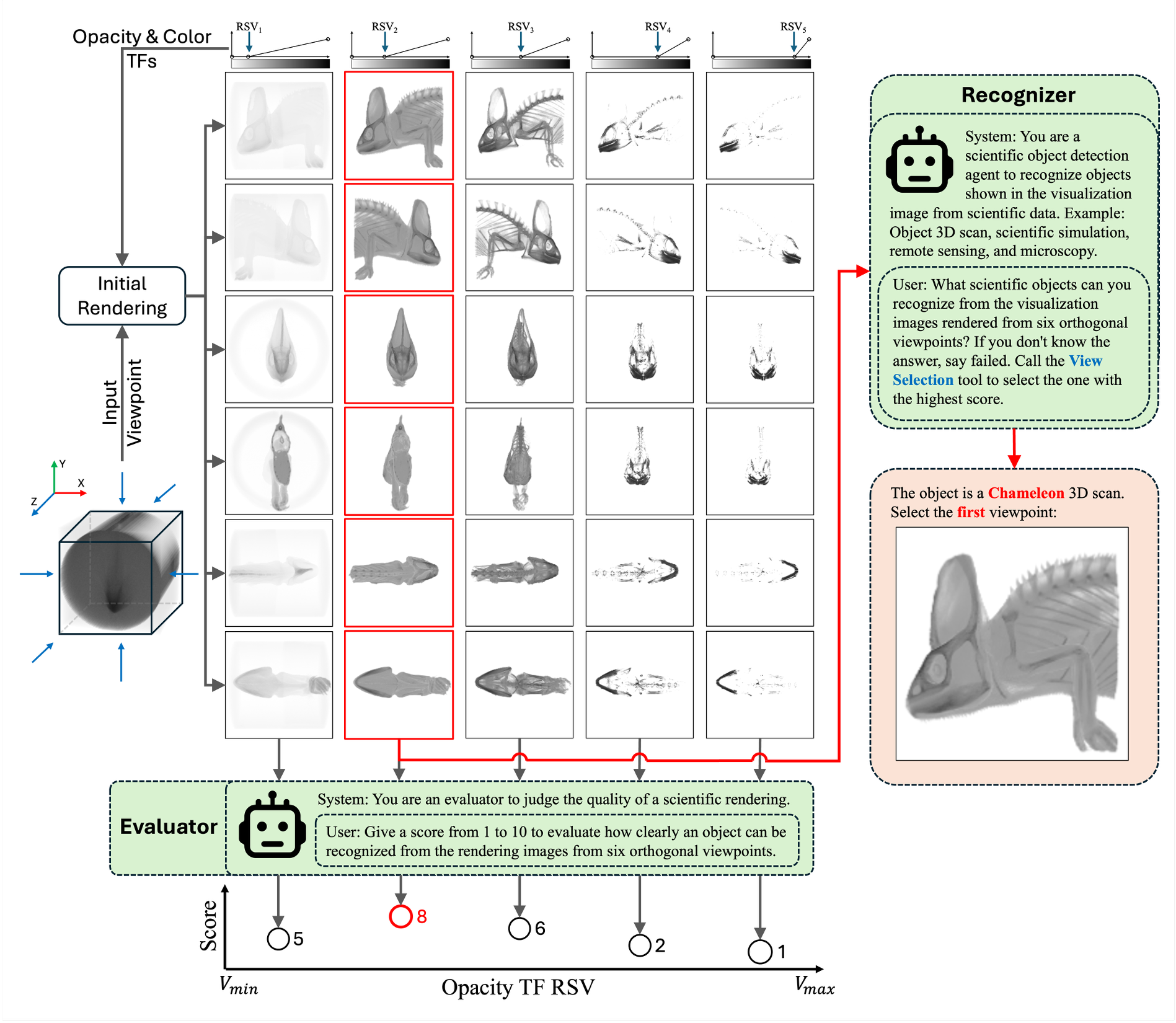}
    \caption{Object recognition agentic workflow through evaluator and recognizer. Five distinct RSVs are used to construct the opacity TFs to render results for the evaluator to judge.}
    \label{fig:recognition}
\end{figure}

\subsection{Data Profiling}
Our agentic workflow starts from data recognition. We leverage the visual understanding and reasoning capability of MLLM to recognize high-level characteristics of the input scientific data.
\subsubsection{Input Profiling}
We first retrieve the metadata out of the raw input data, like the dimension, origin location, spacing, data type (value or label), data structure (regular grid, rectilinear grid, or irregular grid), which will be used to determine the downstream visualization parameters. We also create a downsampled version of the original dataset to enable fast rendering of intermediate visualization inputs for MLLMs throughout the agentic workflow.

\subsubsection{Recognition}
The first agentic step of the workflow is to recognize the intrinsic primary scientific object embedded in the input data. \cref{fig:recognition} shows our agentic workflow for object recognition using the proposed initial rendering, evaluator, and recognizer.

\textbf{Initial Rendering:}
For the initial rendering of the unknown volumetric dataset, we render the downsampled volume using a white-to-black color transfer function in X-ray style. This mapping preserves the original scalar intensity distribution and emphasizes structural gradients while avoiding artificial semantic color cues. Such grayscale renderings highlight geometric boundaries and density variations that are critical features for modern vision-language models, which rely primarily on shape, edge, and intensity patterns for object recognition~\cite{10531703}. Prior work in volume visualization also emphasizes that transfer functions map scalar intensities to optical properties and that intensity-based mappings are sufficient for identifying structures in many datasets~\cite{math12121885}. We select a ramp-shaped opacity transfer function to capture information from the complete range of values. To suppress low-value noise regions and obtain a clearer initial visualization, we introduce a Ramp Starting Value (RSV) that excludes the typical noise range. The ramp opacity transfer function is defined:

\begin{equation}
    Opacity(v)= 
\begin{cases}
    0    & \text{if } v < RSV \\
    \frac{v - RSV}{V_{max} - RSV}  & \text{otherwise}
\end{cases}
,\;\;\;\; RSV \in \ [V_{min}, V_{max})
\end{equation}
where $V_{min}$ and $V_{max}$ are the minimum and maximum values of the input volume. To adapt to arbitrary noise levels across different datasets, we define multiple opacity TFs using different RSVs by evenly sampling $[V_{min}, V_{max}]$. For each opacity TF, we render the downsampled volume from 6 orthogonal viewpoints, each perpendicular to a face of the 3D volume, to obtain a comprehensive set of rendering results for evaluation. Our experiments show that the MLLM can recognize visual representations with rotational invariance, so the up direction of each viewpoint can be chosen from any of the three Cartesian coordinate axes. We use a low resolution for initial renderings to reduce visual embedding token consumption and inference time.

\textbf{Evaluator:}
Our proposed evaluator is an MLLM-as-a-judge that takes the 6 initial rendering results from orthogonal viewpoints and assigns a score from 1 to 10, indicating how clearly an object can be recognized. We call the evaluator for each rendering result using the respective opacity TF, and select the opacity TF that gives the highest score as an optimal TF to generate visualizations for the recognizer.

\textbf{Recognizer:}
The rendering images from the 6 orthogonal viewpoints using the optimal TF are sent to the Recognizer, which is an MLLM that recognizes the objects and suggests one or several keywords for the next knowledge retrieval step. Recognizer can also call the evaluator to automatically select the best viewpoint of the 6 that generates the most representative rendering. The implementation of the view selection workflow is detailed in \cref{sec:view_selection}.

\subsection{Knowledge Retrieval}
\begin{figure}[t]
    \centering
        \includegraphics[trim=0 0 0 0,clip,width=0.9\linewidth]{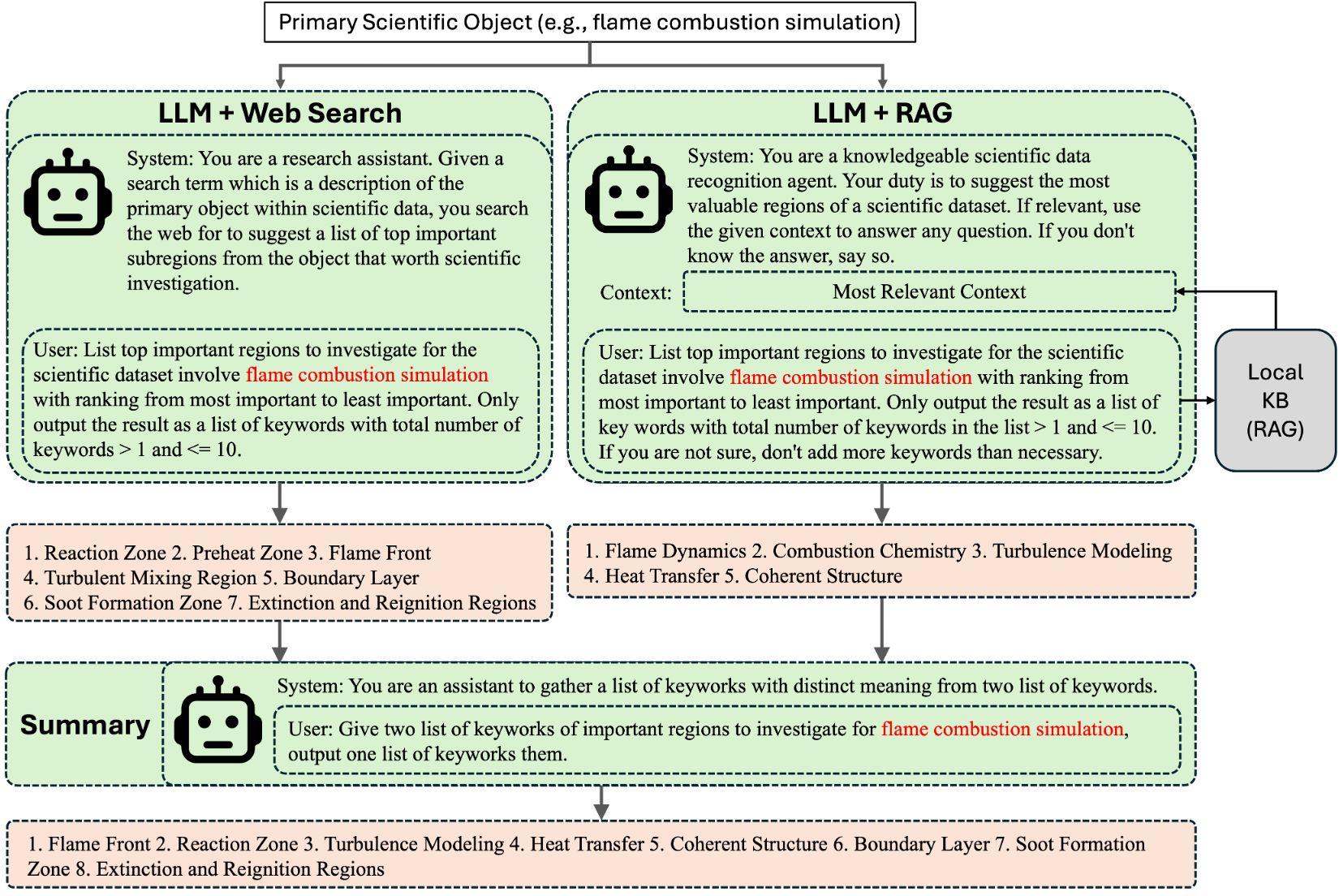}
    \caption{Forager agentic workflow to retrieve knowledge about the regions of interest with scientific significance.}
    \label{fig:kb}
\end{figure}

After identifying the primary scientific object, the agent needs to select regions of interest with the greatest scientific significance. To ensure a comprehensive selection process, we design a forager agent, as shown in \cref{fig:kb}, to first retrieve relevant knowledge from both web searches and a local knowledge base. A summarization LLM then consolidates this information into a set of keywords, which will be used by the downstream steps to guide visualization parameter selection.

\subsubsection{Web Search}
The motivation for incorporating web search into our forager agent is twofold: 1) To leverage search engine ranking mechanisms for retrieving the most relevant and impactful information. 2) To obtain up-to-date results that reflect the latest research trends from diverse sources of scientific visualization. We employ the web search tool from the OpenAI Agents SDK to build the knowledge retrieval agent. Due to the relatively high per-call cost of the web search tool, we adopt lightweight models (e.g., GPT-4o-mini) with a reduced search context size to balance reasoning capability and cost efficiency. \cref{tab:knowledge_retrieval} shows the performance on coverage of the region of interest using various knowledge retrieval methods on flame combustion simulation datasets, LLM with web search gives better coverage than LLM alone.

\begin{table}[h]
  \caption{Regions of interest coverage comparison. Detailed selection of regions can be found in the Appendix.}
  \label{tab:knowledge_retrieval}
  \scriptsize%
	\centering%
  \begin{adjustbox}{width=0.48\textwidth}
      \begin{tabu}{ c | c c c c }
      \toprule
      \multirow{2}{*}{Region of Interest} & LLM  & LLM & LLM & LLM \\
        & Alone & Web Search & RAG & RAG + Web Search \\  
      \midrule
      Flame Front / Stoichiometric Surface    & \textcolor{red}{\ding{55}}       & \textcolor{LimeGreen}{\ding{51}} & \textcolor{LimeGreen}{\ding{51}} & \textcolor{LimeGreen}{\ding{51}}  \\
      \midrule
      Regions Near Flame Surface              & \textcolor{red}{\ding{55}}       & \textcolor{red}{\ding{55}}       & \textcolor{red}{\ding{55}}       & \textcolor{red}{\ding{55}}   \\
      \midrule
      Chemical Reaction Zones                 & \textcolor{LimeGreen}{\ding{51}} & \textcolor{LimeGreen}{\ding{51}} & \textcolor{LimeGreen}{\ding{51}} & \textcolor{LimeGreen}{\ding{51}} \\
      \midrule
      Scalar Dissipation Rate Regions         & \textcolor{red}{\ding{55}}       & \textcolor{red}{\ding{55}}       & \textcolor{red}{\ding{55}}       & \textcolor{red}{\ding{55}}  \\
      \midrule
      Heat Release / High Temperature Regions & \textcolor{red}{\ding{55}}       & \textcolor{red}{\ding{55}}       & \textcolor{LimeGreen}{\ding{51}}       & \textcolor{LimeGreen}{\ding{51}}  \\
      \midrule
      Turbulent Structures / Flow Features    & \textcolor{LimeGreen}{\ding{51}} & \textcolor{LimeGreen}{\ding{51}} & \textcolor{LimeGreen}{\ding{51}} & \textcolor{LimeGreen}{\ding{51}} \\
      \midrule
      Inter-variable Correlation Regions      & \textcolor{red}{\ding{55}}       & \textcolor{red}{\ding{55}}       & \textcolor{red}{\ding{55}} & \textcolor{red}{\ding{55}} \\
      \midrule
      Temporal Coherent Structures            & \textcolor{red}{\ding{55}}       & \textcolor{red}{\ding{55}}       & \textcolor{LimeGreen}{\ding{51}} & \textcolor{LimeGreen}{\ding{51}} \\
      \midrule
      Gradient / Boundary Layers              & \textcolor{red}{\ding{55}}       & \textcolor{LimeGreen}{\ding{51}} & \textcolor{red}{\ding{55}}       & \textcolor{LimeGreen}{\ding{51}} \\
      \midrule
      \end{tabu}
 \end{adjustbox}
\end{table}

\begin{figure*}[t]
    \centering 
    \includegraphics[trim=0 0 0 0,clip,width=\textwidth]{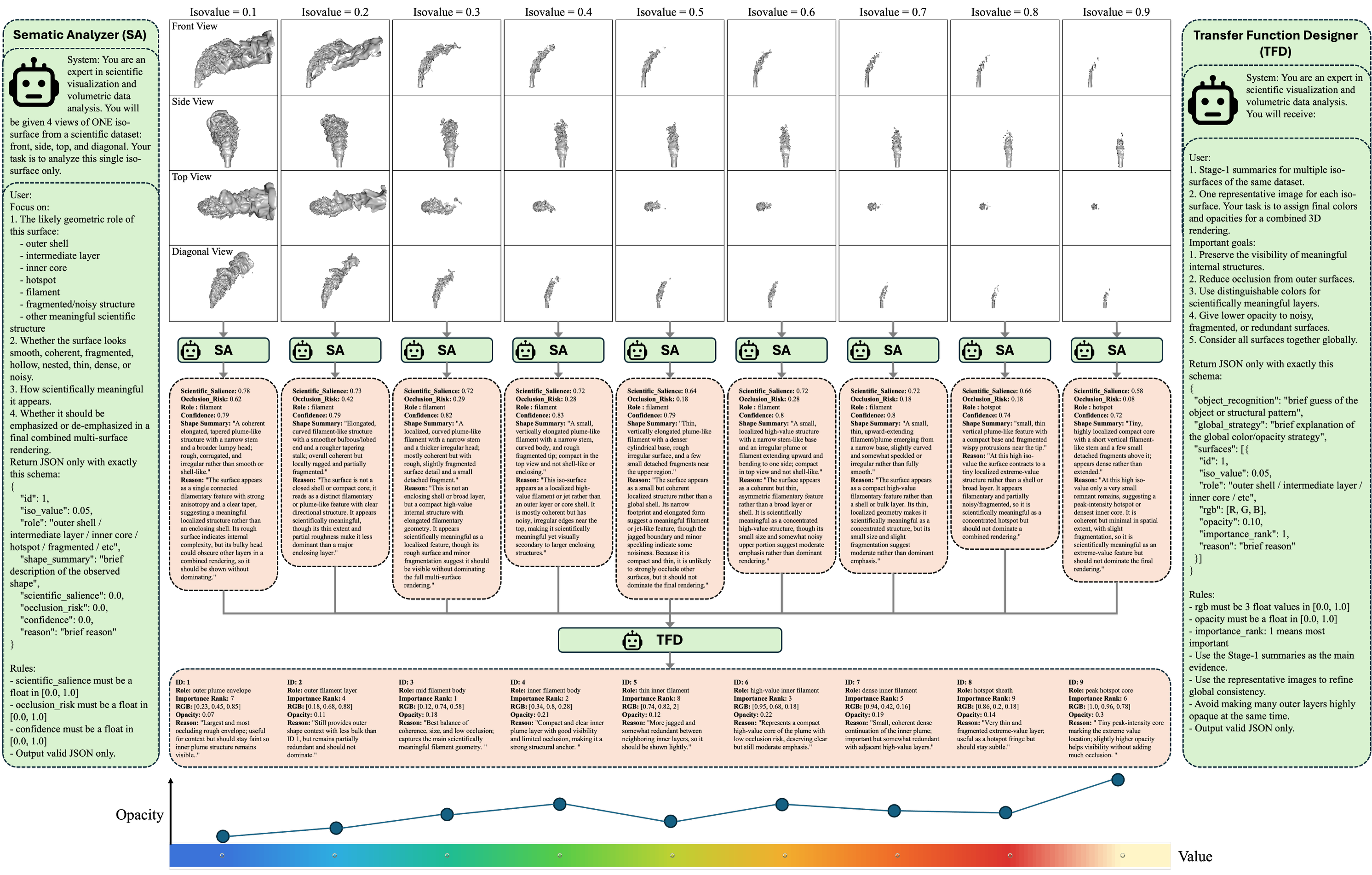}
    \caption{Detailed workflow of Semantic Analyzer (SA) and Transfer Function Designer (TFD) on simulated scientific data (Flame dataset). SA conducts parallel range-of-interest perception for each isovalue, and TFD aggregates the results to generate semantic color and opacity mappings.}
    \label{fig:sa_tfd_simulated}
\end{figure*}

\subsubsection{Knowledge Base}
Although LLM with web search can successfully locate several key regions of interest for the primary scientific object, it is not sufficient for identifying high-quality, domain-specific keywords in a specialized scientific visualization task for the following reasons: 1) Lack of domain-specific depth. 2) Weak alignment with the specific dataset/task. 3) Limited coverage of non-public or niche knowledge. 4) No accumulation of knowledge with stateless web search alone. To enhance the quality of knowledge retrieval, we incorporate a customizable local Knowledge Base (KB) constructed from existing domain-specific research literature tailored to a particular class of scientific data. This KB is fully configurable, allowing the inclusion of diverse textual resources, such as research papers, reference manuals, user guides, and developer tutorials, to construct a comprehensive and searchable repository powered by Retrieval-Augmented Generation (RAG) rather than traditional role-based semantic extraction~\cite{10.1145/3544548.3581067}.

When building the RAG, we first convert the textual document from its respective format into Markdown for its lightweight structure and strong alignment with the representations used in modern LLM foundational models. Then we vectorize the chunks of the entire markdown document into a vector database. We select LangChain~\cite{langchain2023} to build the RAG for its flexibility and scalability. The auto-encoding LLM or the embedding model we use to convert textual chunks into vectors is the text-embedding-3-large Model from OpenAI. Compared to other text embedding models like all-MiniLM-L6-v2 from Hugging Face Sentence-Transformers models and Google Gemini-embedding-001, text-embedding-3-large has superior performance on semantic quality, consistency, and long input support~\cite{goel2025sage, ranjan2025one}. When we call the LLM for knowledge retrieval using the RAG, we set a small temperature argument (0.1) for a stronger precision and deterministic guarantee for reliability. We use the open-source Chroma as the vector database and select GPT-4.1-nano for this task due to its low cost for instruction-based calls and fast response time. By default, the proposed SASAV operates effectively by relying solely on LLMs with web search for knowledge retrieval to enable a fully autonomous agentic workflow. The RAG local knowledge base is not mandatory, but is recommended for the forager agent to retrieve a more comprehensive and domain-specific knowledge. \cref{tab:knowledge_retrieval} shows that LLM with both RAG and web search gives the best coverage for the region of interest.

\begin{figure*}[t]
    \centering 
    \includegraphics[trim=0 0 0 0,clip,width=\textwidth]{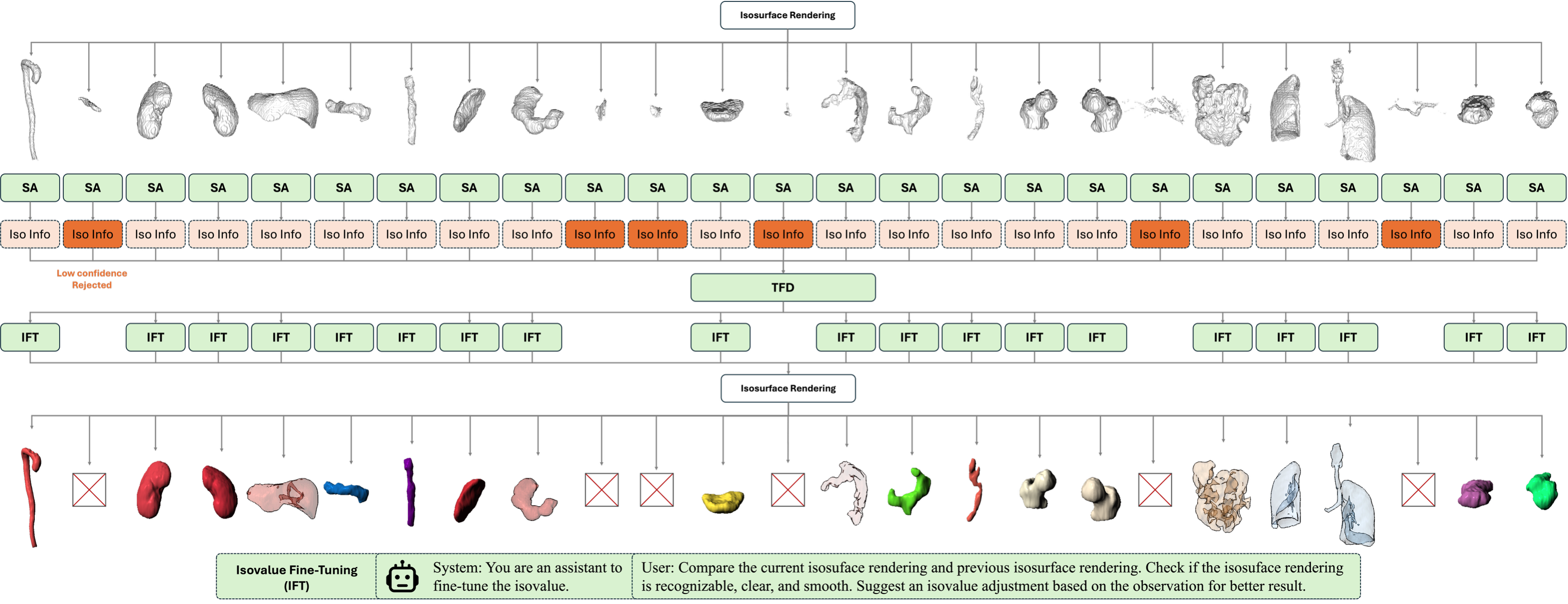}
    \caption{Example workflow color and opacity mapping suggestion on empirical scientific data (AbdomenAtlas 1.0 Mini dataset). The SA output shown in dark orange indicates a failure to recognize the isosurface due to very low confidence score, consequently, the corresponding isosurface is excluded from the final rendering by the TFD. IFT adjusts the isovalue to produce smoother and more clearly defined organ surfaces.}
    \label{fig:sa_tfd_empirical}
\end{figure*}

\subsection{Transfer Function Suggestion}
Based on the type of scientific object, simulated or empirical, SASAV is designed to recommend appropriate color and opacity mappings for the final visualization. For simulated data, we employ direct volume rendering (DVR) for the final visualization, whereas for empirical data, we use isosurface rendering. This is because a simulated scientific object does not exhibit distinct boundaries, due to its continuous value distribution and strong spatial correlation throughout the volume; consequently, the full value spectrum is meaningful and should be represented in the final visualization. So we leverage the continuous color and opacity transfer functions to generate the DVR. An empirical scientific object is a reconstruction of a real-world entity with rich semantic meaning, characterized by clearly defined boundaries between distinct, spatially separated subregions. To highlight structural shapes and their relationships, we employ discrete color and opacity mapping to present these subregions via isosurface rendering. Recent studies indicate that multimodal foundation models, such as vision–language models, can support visualization design by analyzing visualization images, evaluating their effectiveness with respect to user goals, and providing improvement suggestions~\cite{10756172, 11075548}. Existing MLLM-based optimization on TFs relies on a single-threaded iterative process through a feedback loop, giving slow convergence and limited generalizability~\cite{https://doi.org/10.1111/cgf.15093, wang2025intuiTF, 10771134}. To achieve efficient and scalable transfer function optimization for large-scale scientific data, we propose a swarming then consolidating design~\cite{11153807} that employs parallel multi-agent MLLM analyzers to reason, perceive, and summarize local features (\cref{sec:rir}), followed by a global designer that synthesizes these insights to derive optimal TFs (\cref{sec:scom}).

\subsubsection{Range of Interest Perception}\label{sec:rir}
To highlight the most informative regions with the greatest scientific significance, a range of interest must be derived from the value spectrum of the input scalar field. We perform a range of interest selection by leveraging the visual reasoning capabilities of frontier MLLMs, as demonstrated in recent studies~\cite{11075548, 10756172} and validated by benchmarks such as ARC-AGI-2~\cite{lu2024mathvista, chollet2025arc} and MMMU-Pro~\cite{yue-etal-2025-mmmu}. To capture a complete set of regions of interest, we evaluate the full value range of the input volumetric scalar field, enabling a global perception of the dataset for more comprehensive visualization. We first perform uniform sampling across the value range of the input volume. If the input data contains a limited set of discrete integer values, such as label data, we sample each value directly. For input data with a continuous value spectrum, the sample distance $d$ is defined as:
\begin{equation}
    d=(V_{max} - V_{min})/(M + 1)
\end{equation}
where $V_{max}$ and $V_{min}$ are the upper and lower bounds of the input volumetric scalar field. $M$ is the number of value samples within $(V_{min}, V_{max})$. For each sampled value, we generate a group of isosurface renderings from four viewpoints: front, side, top, and diagonal. To facilitate visual perception by the MLLM, we assign a white color to the isosurfaces and enable shading to convey depth information against a white background.

We design a Semantic Analyzer (SA) to evaluate each isosurface rendering group based on key visualization metrics, as shown in the left-hand side of \cref{fig:sa_tfd_simulated}. We design the prompt to guide SA to focus on critical aspects of visualization quality: 1) Geometric role identification to capture high-level structure; 2) Surface evaluation to characterize the isosurface rendering; 3) Scientific significance to assess the meaningfulness of the isosurface; 4) Importance estimation to determine its contribution to the final global rendering. We also frame the SA output to provide both quantitative evaluations and textual descriptions for each isosurface rendering. Unlike prior work that formulates MLLM outputs as binary judgments (e.g., recognizable vs. not, clear vs. unclear)\cite{https://doi.org/10.1111/cgf.15093}, our approach provides fine-grained evaluation through numerical scores on a 10-point scale. The numerical metrics include: 1) A scientific salience score to assess the degree to which the isosurface conveys relevant scientific information; 2) An occlusion risk score to measure the likelihood of visual occlusion; 3) A confidence score to estimate the model’s certainty in its recognition. The textual outputs include: 1) A shape summary describing the observed geometry; 2) A brief explanation justifying the derived metrics. We constrain the range of the numerical metrics and output format at the end. SA operates as a swarm agent to process isosurface renderings in parallel for efficient MLLM inference.




\subsubsection{Semantic Color and Opacity Mapping}\label{sec:scom}
We propose the Transfer Function Designer (TFD), which leverages the isovalue-dependent insights produced by SA to recommend corresponding color and opacity mappings for respective isovalue, as shown in the right-hand side of \cref{fig:sa_tfd_simulated}. Since we adopt different visualization methods, DVR for simulated data and isosurface rendering for empirical data, TFD needs to output different types of color mapping. For simulated data, a continuous color and opacity transfer function is generated by linearly interpolating the suggested color and opacity across the isovalues, which will be used for rendering the final visualization. For empirical data, there are two extra steps needed before rendering the final visualization, as shown in \cref{fig:sa_tfd_empirical}. First, an isovalue rejection is performed by the TFD to discard isovalues assigned low confidence by SA, indicating that no meaningful semantic structure is discernible in the corresponding isosurface rendering. Rejected isovalues are excluded from the final rendering by setting their opacity to zero. Second, for all the meaningful isovalues, we propose Isovalue Fine-Tuning (IFT) to update the color and opacity mapping. Since the initial isovalues are uniformly sampled, IFT refines each isovalue to better align with the true surfaces of the underlying empirical structures. For the $i$ th isovalue with value of $v_i$, we want to find the optimal isovalue $v_{opt}$ within its neighboring range that maximizes the recognition score on recognizability, clarity, and smoothness.

\begin{equation}
v_{opt} = \arg \max_{v}Score(v), v \in [v_{i - 1}, v_{i + 1}] 
\end{equation}

IFT accomplishes this by leveraging the comparative capability of MLLMs, which are provided with a prior rendering at one isovalue and a new rendering at another. The MLLM then recommends adjustment of the isovalue toward the one that produces a higher recognition score.

\begin{figure*}[t]
    \centering 
    \includegraphics[trim=2 2 2 2,clip,width=\textwidth]{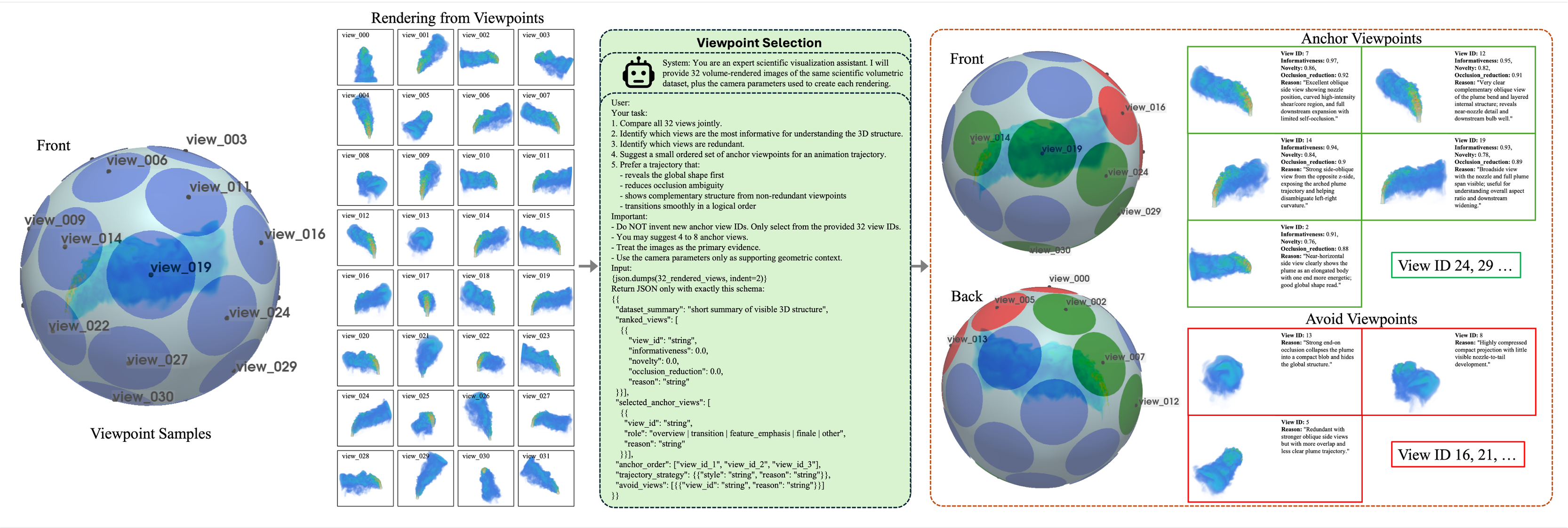}
    \caption{View selection process is to recommend the anchor views, the most informative views, and avoid views that are redundant or have occlusion. This example shows 32 viewpoints sampled on the view sphere through a Fibonacci lattice. Anchor viewpoints are marked in green, while the avoid viewpoints are marked in red.}
    \label{fig:view_selection}
\end{figure*}

\subsection{View Selection}\label{sec:view_selection}
SASAV introduces a view evaluation workflow that selects the most informative viewpoint for visualization and recommends an exploratory trajectory to support more comprehensive data exploration through animation. The criteria for selecting an effective viewpoint in scientific visualization include: 1) Covering the majority of regions of interest, 2) Minimizing attenuation and occlusion, and 3) Maximizing the visibility of value intervals of interest. The View Selection (VS) step of SASAV leverages the visual evaluation capability of MLLM on multiple visual inputs. We adopt the concept of view sphere~\cite{1532833}, defined as a sphere enclosing the dataset with its centroid aligned to the origin of the data volume. First, a predefined number ($K$) of candidate viewpoints is evenly sampled from the view sphere using a Fibonacci lattice. Second, a visualization image is rendered from each sampled viewpoint using the color and opacity mappings derived from the previous TFs suggestion step. Third, all $K$ candidate viewpoints are provided to the MLLM to identify informative viewpoints (anchor viewpoints) and those to avoid. \cref{fig:view_selection} demonstrates the view section process. We design the prompt to guide VS in the following tasks: 1) Performing a global comparison of renderings from all $K$ viewpoints; 2) Identifying the most informative viewpoints; 3) Detecting redundant viewpoints; 4) Selecting anchor viewpoints for constructing an exploratory trajectory; 5) Incorporating preferred characteristics of the trajectory. We also frame the output to suggest 3 lists of viewpoints: 1) Viewpoints ranked by the informativeness score; 2) Anchor viewpoints with the order for trajectory construction; 3) Viewpoints to avoid. It is worth noticing that there is a trade-off among the granularity of view selection, input image resolution, and MLLM performance. Increasing the number of viewpoint candidates $K$ will improve the accuracy of best view selection. However, feeding too many images to MLLM has the following issues: 1) Visual embedding converts input images into tokens determined by their resolution. Supplying many high-resolution viewpoint renderings can significantly increase token consumption. 2) Providing too many input images can lead to a compressed attention effect, where the model focuses on only a few images while ignoring others, resulting in unstable outputs and diminished ranking reliability. In this work, we select $K=32$ and constrain the viewpoint rendering resolution as $256\times 256$ to balance the accuracy and performance. SASAV creates a dense exploratory trajectory from the anchor viewpoints through high-order Catmull–Rom splines interpolation for animation generation.



\subsection{Implementation}
\subsubsection{Prompting techniques}
To enhance the reasoning capability and prediction stability of SASAV, we adopt the following prompting techniques:

\begin{itemize}[leftmargin=*]
  \item When querying the MLLM, we use a low temperature (0.1) to promote stable and consistent outputs. This is particularly important for the strict JSON formatting required by SASAV, as lower temperatures produce more predictable results across trials.
  \item When designing prompts, we incorporate in-context learning with few-shot examples. For instance, in SA’s geometric role suggestion, we include predefined options (e.g., outer shell, intermediate layer, inner core, hotspot, filament, fragmented/noisy structure, and other meaningful scientific patterns) within the prompt to improve the quality of the responses.
  \item Our prompt design leverages inference-time reasoning by incorporating richer contextual information, thereby enhancing reasoning capability in line with the Chinchilla scaling law. Given the diminishing returns of training-time optimization, SASAV prioritizes inference-time strategies rather than fine-tuning an open-source MLLM model.
  \item Given the absence of prior domain knowledge, SASAV adopts intent-driven prompting rather than specification-driven prompting in its task descriptions, allowing it to better leverage the reasoning capabilities of MLLMs.
  \item We decouple the control from tools to agents. So tools perform only predefined and hardcoded tasks, while agents maintain and dynamically adjust all control parameters as needed.
  \item When querying the MLLM, we employ levels of prompts, system prompt, and user prompt, to prevent prompt injection.
  \item SASAV also leverages asynchronous I/O to call multiple concurrent MLLMs. 
  \item To reduce LLM hallucinations, we append the instruction “If you don’t know the answer, say ‘failed.’” to the end of the user prompt.
\end{itemize}

\subsubsection{Visualization Generation}

Visualization generation is the final stage of SASAV, where images and animations are produced from the input scientific data using the transfer functions and viewpoints suggested by the agent. SASAV is implemented as a self-contained system for visualization generation and interaction, with optimizations including data downsampling, multi-threading, and GPU acceleration. It supports routine exploration and moderate-scale rendering on local resources, while high-resolution rendering of large-scale datasets can be accelerated through distributed volume visualization on high-performance computing clusters~\cite{10386434, 291532}. Throughout the workflow, the visualization modules are implemented in C++ using VTK to ensure efficiency, flexibility, and stability. We do not rely on recent visualization code-generation agents~\cite{11298870}, as their Pass@1 performance remains inconsistent and they may introduce variability into the generated results.

\subsubsection{Interactive Visualization}
In addition to generating visualization images and animations, SASAV provides an interactive interface that enables users to explore the data in real time. SASAV also stores the recommended visualization parameters, like viewpoints, color, and opacity mappings, in JSON and color table (.ct) format, enabling seamless import into interactive scientific visualization tools such as ParaView and VisIt.

\subsubsection{User Interface}
\begin{figure}[t]
    \centering
        \includegraphics[trim=2 2 2 2,clip,width=\linewidth]{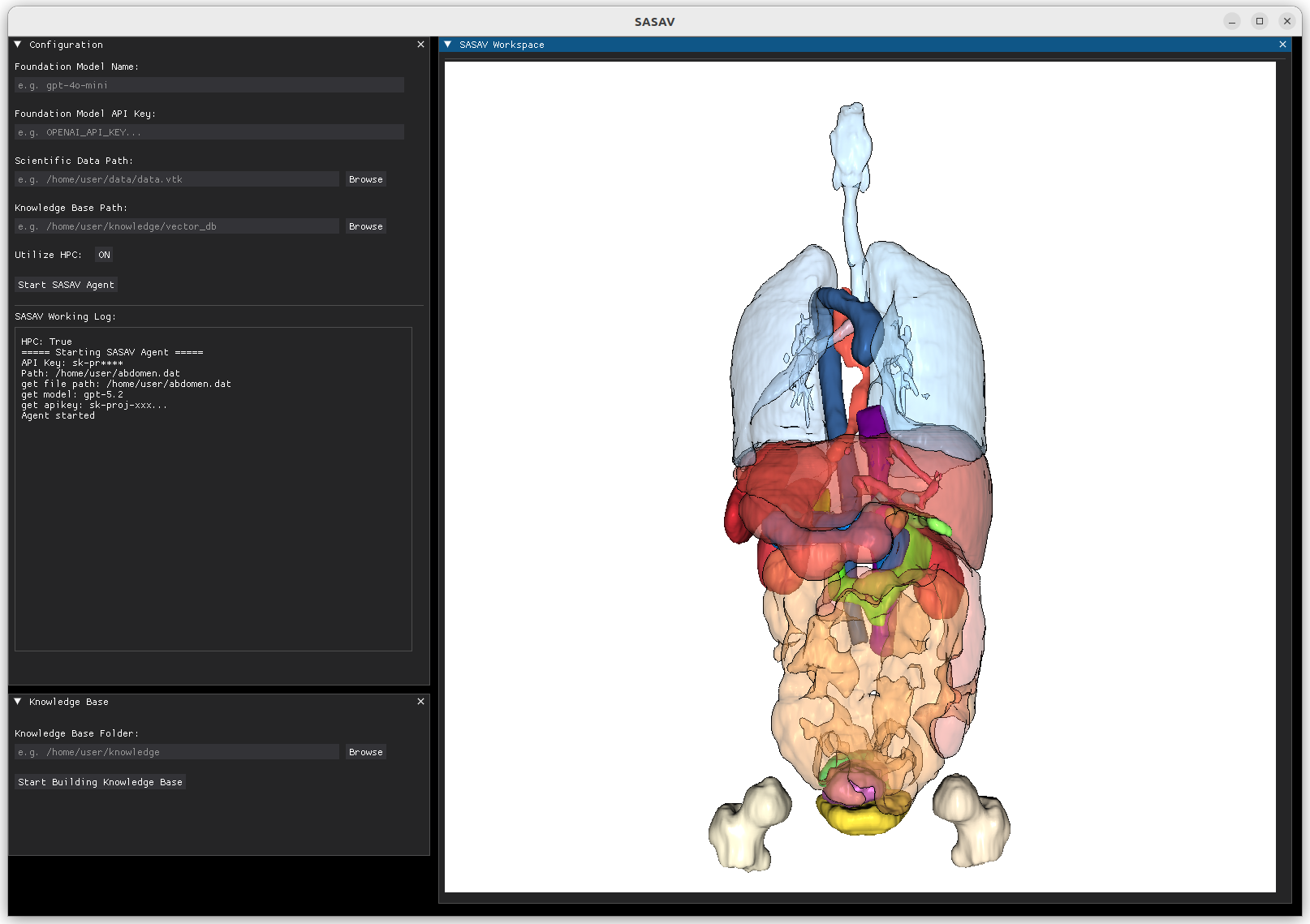}
    \caption{User interface of SASAV.}
    \label{fig:ui}
\end{figure}

The SASAV interface is intuitive and minimal, helping users visualize their own scientific data with ease. We followed the design patterns suggested for future Agentic Visualization (AV)~\cite{11153807} to construct the SASAV UI. As shown in \cref{fig:ui}, the interface consists of three main components: configuration panel, knowledge base panel, and workspace panel. The configuration panel holds basic configurations, including foundation model selection, API key, input scientific data selection, knowledge base selection, and HPC rendering option. The user can click the "Start SASAV Agent" button to initiate the agentic workflow. It also contains a text window to display the working log while the SASAV agent is operating. The knowledge base panel is for building a customized RAG-based vector database from user-specific documents. The workspace panel presents intermediate visualization outputs in real-time as the agentic workflow progresses.

\section{Results and Evaluation}
\subsection{Experimental Setup}
For all MLLMs used in our agentic workflow, we employ OpenAI's reasoning model GPT 5.4 due to its overall performance in visual reasoning, performance, and cost at the time of writing. We implement the agentic workflow of SASAV using OpenAI Agents SDK. JSON format is used to represent intermediate textual outputs, enabling structured data exchange across different MLLMs throughout the pipeline. For the hyperparameter of SASAV, we set the number of RSV $N=5$, the number of isovalues $M=9$, and the number of sampled viewpoints $K=32$. When constructing the local knowledge base, we use LangChain to process the collection of Markdown documents, converted from diverse source formats, by segmenting them into chunks of size 1000 with an overlap of 200, improving contextual coherence and boundary continuity, particularly for research literature.  We also force both web search and RAG MLLMs to output regions of interest in the order of importance and set an upper bound of 10 on the number of keywords. We downsample the input volume within a cube with a resolution of $256\times 256\times 256$ and normalize its value to $[0, 1]$. All the intermediate rendering resolutions are set to $256\times 256$. 

\subsection{Datasets}
We evaluate SASAV across 5 diverse scientific datasets to demonstrate its performance and generalizability. As detailed in \cref{tab:datasets}, AbdomenAtlas 1.0 Mini and Chameleon are empirical object types of data from 3D scanning, while Miranda, Flame, and Richtmyer are simulated object types of data from scientific simulations. The volume size ranges from 260 MB to 7.5 GB.

\begin{table}[t]
  \caption{Scientific volumetric datasets used in the experiments.}
  \label{tab:datasets}
  \scriptsize%
	\centering%
  \begin{adjustbox}{width=0.48\textwidth}
      \begin{tabu}{ c c c c c }
          \toprule
          Dataset & Scientific Object Type & Resolution & Data Type & Size \\
          \midrule
          Abdomen (AbdomenAtlas 1.0 Mini)  & Empirical  & $511\times 404\times 339$ & float32 & 0.26 GB\\
          \midrule
          Chameleon             & Empirical  & $1024^2 \times 1080$ & uint16 & 2.11 GB\\
          \midrule
          Miranda               & Simulated  & $1024^3$ & float32 &  4.0 GB \\
          \midrule
          Flame                 & Simulated  & $1408 \times 1080 \times 1100$ & float32 &  6.23 GB \\
          \midrule
          Richtmyer             & Simulated  & $2048^2 \times 1920$ & uint8 &  7.5 GB \\
          \bottomrule
      \end{tabu}
  \end{adjustbox}
\end{table}

\begin{figure*}[t]
    \centering 
    \includegraphics[trim=2 2 2 4,clip,width=0.94\textwidth]{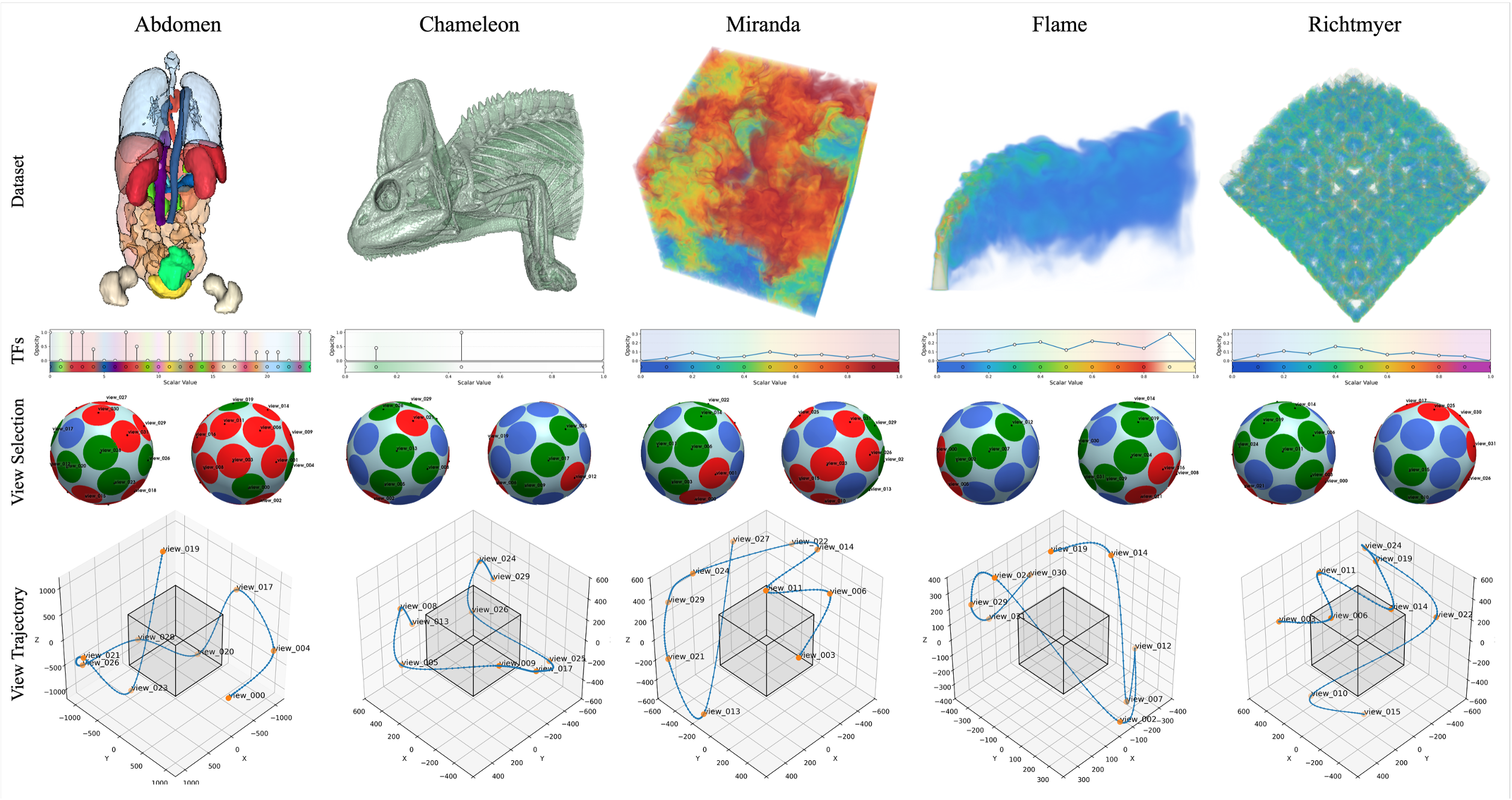}
    \caption{Final visualization image generated by SASAV and its suggested visualization parameters of TFs, anchor viewpoints, and exploratory trajectory for all the datasets considered.}
    \label{fig:all_data}
\end{figure*}

\subsection{Evaluation}
\subsubsection{Quality}
Since there is no single ground truth to quantify the rendering quality of generated visualization from an agentic workflow, we perform a qualitative evaluation. \cref{fig:all_data} shows the final visualization image generated by SASAV for all the datasets, together with suggested intermediate visualization parameters, such as color and opacity mapping (for Abdomen and Chameleon datasets), TFs (for Miranda, Flame, and Richtmyer datasets), selected anchor viewpoints, avoid viewpoints, and exploratory trajectory. For empirical object types, such as the Abdomen and Chameleon datasets, SASAV effectively highlights key regions of interest with realistic color, for instance, distinct organs within the abdomen and the skin and skeletal structures of the chameleon. For simulated object types, such as the Miranda, Flame, and Richtmyer datasets, SASAV effectively visualizes value distributions using distinct color mappings, thereby revealing the intrinsic structures embedded within the volume. Combined with animation and an interactive UI, SASAV delivers highly informative visualizations directly from the data, without requiring prior domain knowledge or cumbersome iterative exploration through user feedback.

\subsubsection{Performance}
\begin{figure}[t]
    \centering
        \includegraphics[trim=2 2 2 2,clip,width=\linewidth]{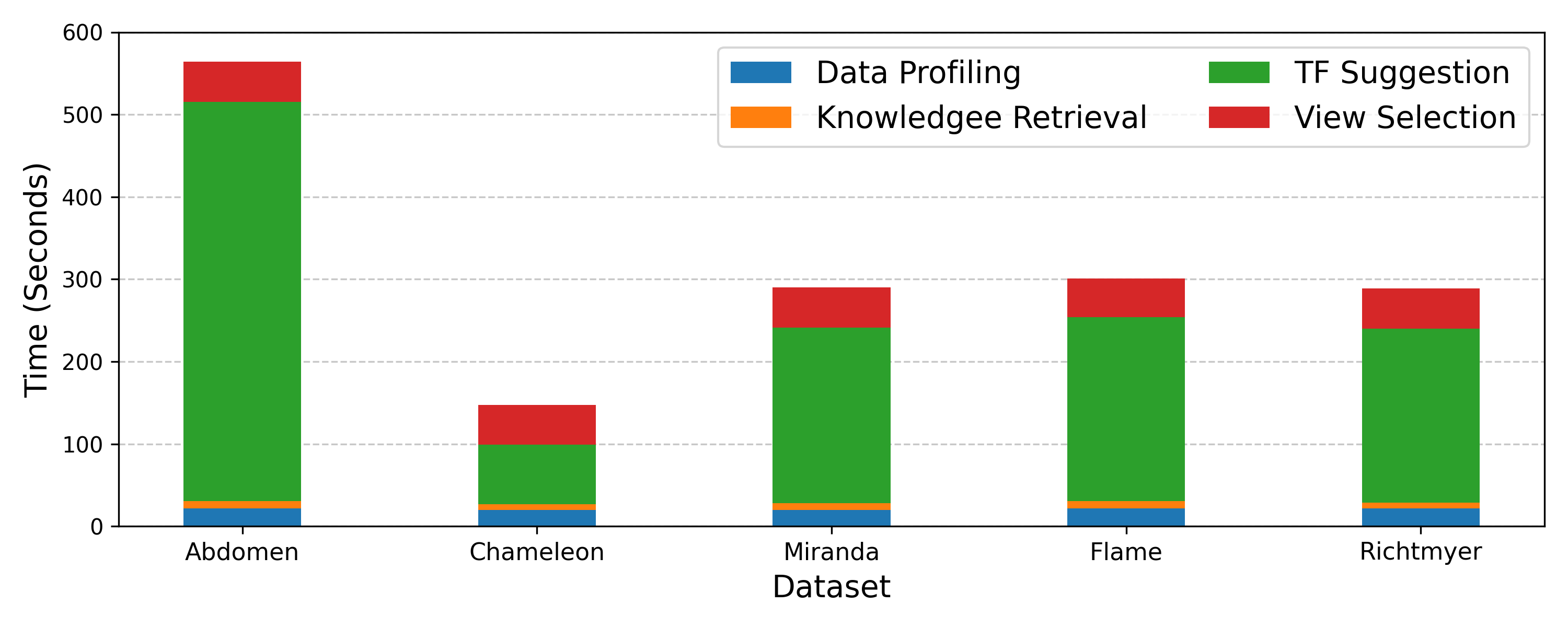}
    \caption{Time consumption of SASAV for each step across all 5 datasets. Time duration reported as the average of five repeated trials. }
    \label{fig:time}
\end{figure}

We measure the time consumption of each key step of the SASAV workflow and evaluate the efficiency of the proposed agentic workflow. \cref{fig:time} shows the time breakdown across the key steps of data profiling, knowledge retrieval, TF suggestion, and view selection for all the datasets. Among all the steps, TF suggestion takes the longest time due to two reasons: First, the Semantic Analyzer (SA) takes multiple images as input, which results in a large amount of input tokens, and its output is also required to be comprehensive. Second, the Transfer Function Designer (TFD) needs to conclude optimal color and opacity mapping from the output of SA, which requires heavy reasoning. Moreover, TF suggestion is also correlated with the number of meaningful isovalues, for it determines the number of images sent to the SA. Since the simulated type datasets (Miranda, Flame, and Richtmyer) evenly sample the same number of isovalues, they have similar time spent on the TF suggestion step. However, for empirical type datasets (Abdomen and Chameleon), their meaningful isovalue is data-dependent, giving very different time consumption of TF suggestion. Compared with the Chameleon dataset, the Abdomen dataset has more meaningful isovalues, and each represents one of the many organs, resulting in the highest time consumption in the TF suggestion step. We also measured the time required to render visualization images and animations, as reported in \cref{tab:rendering_time}. The results indicate that local execution is adequate for routine rendering, whereas high-performance computing substantially reduces latency for large-scale and animation-intensive workloads.
 
\subsubsection{Token Usage}
\begin{figure}[t]
    \centering
        \includegraphics[trim=2 2 2 2,clip,width=\linewidth]{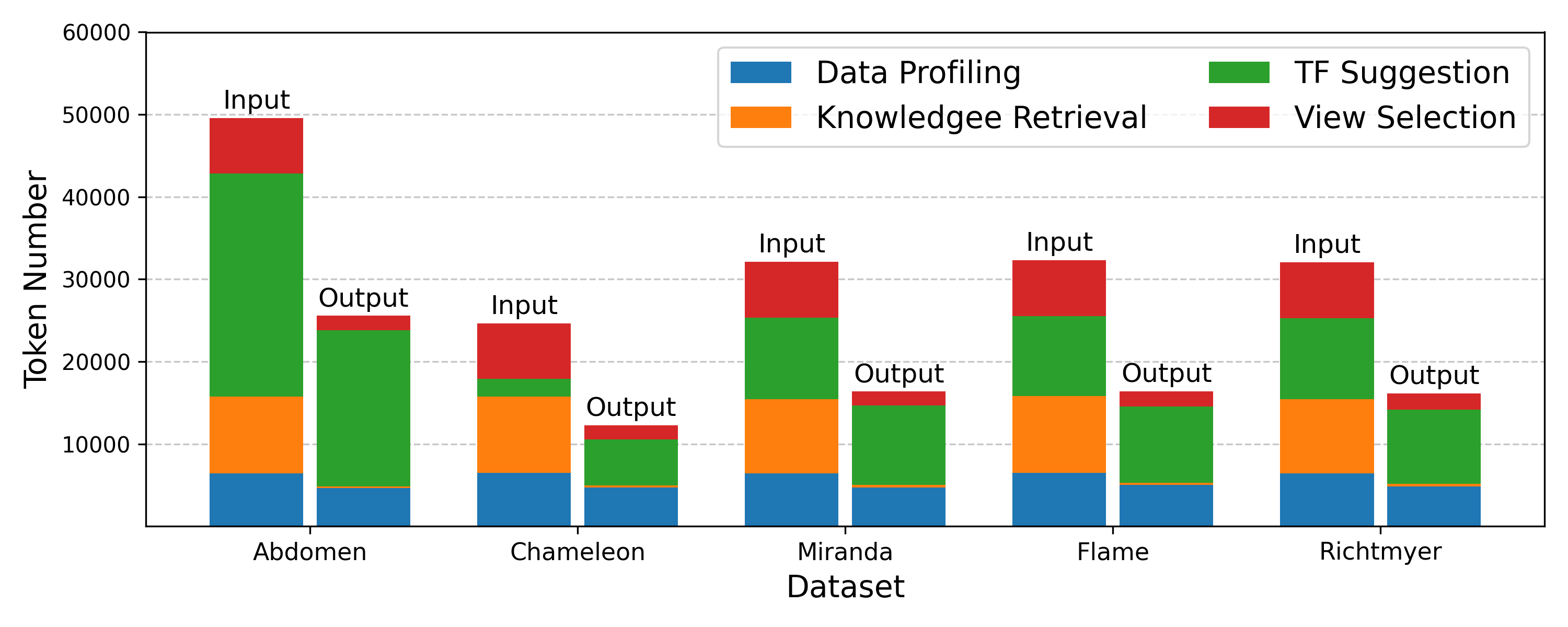}
    \caption{Token usage of SASAV for each step across all 5 datasets. Input and output tokens are reported as average of five repeated trials. }
    \label{fig:tokens}
\end{figure}

We record both the input and output token usage for each key step of SASAV while processing the scientific datasets, as shown in \cref{fig:tokens}. We have several observations: 1) Input token usage exceeds output tokens because many queries throughout the workflow incorporate image inputs, which produce a large number of tokens after embedding. 2) The number of tokens used during the TF suggestion step for the empirical object type dataset is related to the number of input isosurfaces, which determines the number of input images to MLLM. 3) The view selection step also takes images from different viewpoints as input, giving high token usage as well.

\begin{table}[t]
  \caption{High resolution ($2048\times 2048$) rendering for image and animation on local machine (8 cores) and HPC leveraging 128 cores.}
  \label{tab:rendering_time}
  \scriptsize%
	\centering%
  \begin{adjustbox}{width=0.49\textwidth}
      \begin{tabu}{ c | c | c | c | c | c | c }
      \toprule
      \multirow{2}{*}{Dataset} & \multicolumn{2}{c|}{Image (Seconds)} & \multicolumn{4}{c}{Animation (Hours)} \\
        & 
      Local $\downarrow$  & HPC (128) $\downarrow$  &
      Anchor Viewpoints Num & Trajectory Size & Local $\downarrow$ & HPC (128) $\downarrow$\\
      \midrule
      Abdomen & 42.3  & 1.5 & 9  & 960 & 8.7 & 0.31\\
      \midrule
      Chameleon &  15.3  & 1.1 & 9  & 960 & 2.9 & 0.29 \\
      \midrule
      Miranda & 37.8 & 1.2 & 10 & 1080 & 9.2 & 0.31\\
       \midrule
      Flame & 41.6  & 1.4 & 9  & 960 & 9.0 & 0.35\\
      \midrule
      Richtmyer & Failed & 3.8 & 9  & 960 & Failed & 0.98\\
      \bottomrule
      \end{tabu}
 \end{adjustbox}
\end{table}

\subsection{Experts Feedback}
In addition to qualitative evaluation, we also collected expert feedback from the scientific visualization community. Specifically, we consulted three researchers in scientific data analysis and visualization: a professor specializing in 3D spatial data, a researcher focused on large-scale data visualization, and a senior Ph.D. student working on volume visualization. They all have experience in using existing scientific visualization tools like ParaView. We introduced SASAV to the experts and asked them to independently run the agent using the same set of datasets used for the experiment in the paper. 

\textbf{The professor said:} \textit{"This system is highly effective in reducing manual effort in volume rendering. Its primary advantage lies in its efficiency, significantly saving both time and human labor."} The professor also noted that SASAV has strong potential to evolve into a more closed-loop framework by incorporating additional agents for downstream tasks that evaluate and refine the visualization results.

\textbf{The researcher said:} \textit{"The automatic visualization parameter searching can save a lot of time, especially when the dataset is large. It is also convenient to be able to use HPC through SASAV."} The researcher is curious about introducing more visualization techniques for more types of data, for example, point clouds and time-varying data.

\textbf{The Ph.D. student said:} \textit{"Scientists rely less on technical knowledge to explore their data, which can help speed up analysis time. If it is light, it can possibly be deployed on smaller or weaker machines, which is convenient if you do not want to do some quick analysis on the fly."} The Ph.D. student also highlighted challenges that arise when multiple scientists collaborate on the same dataset. A fully autonomous agent can support such collaboration by offering additional perspectives on the shared research problem.

\section{Limitations and Future Work}
Our work is the initial effort toward full autonomous data analysis to accelerate scientific discovery through agentic workflows. Due to the inherent randomness of LLMs, the resulting visualizations may vary across runs, particularly in terms of color. For color selection, our agent performs better on empirical objects than on simulated objects, as real-world objects carry richer semantic meaning that aligns more effectively with MLLMs. Additionally, the proposed knowledge base (KB) can be further enhanced by incorporating a broader range of scientific analysis use cases, such as insights from future research in scientific data understanding or emerging benchmarks for scientific visualization agents~\cite{ai2025evaluation}. In this paper, we focus exclusively on Direct Volume Rendering (DVR) and isosurface visualization. Other techniques, such as cutaway views, slice views, and feature-based visualization methods, can be further explored in the future. The workflow of SASAV can be extended to handle other types of scientific data, such as multivariate volume, tensor field, and time-varying data. The performance of SASAV can be further optimized by customizing the selection of different MLLMs to fully utilize their respective strengths, achieving better results with minimal cost. SASAV currently relies on vision large language models as the evaluator. It would be valuable to explore the use of video vision transformers (ViViT)~\cite{Arnab_2021_ICCV, 10982110} as an alternative, enabling more direct understanding of time-varying scientific data. SASAV can also be improved by leveraging middleware, like ACADEMY~\cite{pauloski2025empowering}, to deploy autonomous agents directly across the federated research ecosystem, including HPC systems, experimental facilities, and data repositories.

\section{Conclusion}
In this work, we present SASAV, the first fully autonomous self-directed agent for scientific analysis and visualization. SASAV proposes an agentic workflow capable of understanding the data, highlighting scientific significance, and suggesting visualization parameters without prior knowledge or human-in-the-loop, enabling a comprehensive, systematic, and scalable visualization pipeline. The performance of SASAV is expected to improve naturally as frontier LLMs continue to advance. We conduct a comprehensive evaluation of SASAV across visualization quality, execution efficiency, and token cost, highlighting its potential as a foundational building block for AI-driven science that accelerates discovery and innovation.


\bibliographystyle{abbrv-doi-hyperref}

\bibliography{template}

\end{document}